\documentclass[prd,twocolumn,superscriptaddress,showpacs,preprintnumbers,amsmath,amssymb,APS]{revtex4-1}
\usepackage{bm}
\usepackage[latin1]{inputenc}
\usepackage[spanish,english]{babel}
\usepackage{amsfonts}
\usepackage{amssymb}
\usepackage{graphicx}
\usepackage{color}

\newcommand{\be}{\begin{equation}}
\newcommand{\ee}{\end{equation}}
\newcommand{\bea}{\begin{eqnarray}}
\newcommand{\eea}{\end{eqnarray}}

\begin{document}

\title{{\bf Fifty years of 
 cosmological particle creation}}

\author{Leonard Parker}\email{leonard@uwm.edu}\affiliation{Physics Department, University of Wisconsin-Milwaukee,
Kenwood Integrated Research Complex, 3135 North Maryland Avenue, Milwaukee, WI 53211, USA.}
\author{José Navarro-Salas}\email{jnavarro@ific.uv.es}
\affiliation{Departamento de Fisica Teorica and IFIC, Centro Mixto Universidad de Valencia-CSIC. Facultad de Fisica, Universidad de Valencia, Burjassot-46100, Valencia, Spain.}

\date{\today}

\begin{abstract}

In the early sixties Leonard Parker discovered that the expansion of the universe  can create particles out of the vacuum, opening a new and fruitfull field in physics. 
We give a historical review in the form of an interview that took place  during the Conference ERE2014 (Valencia 1-5, September, 2014).

\end{abstract}


\maketitle

\section{Preamble}

Prof. Parker is the pioneer of the theory of quantized fields in curved spacetime. His breakthrough discovery, in the early sixties,  that the expansion of the universe  can create particles out of the vacuum opened a new field in physics, with many treasures inside. 
This surprising result was stated and analyzed with detail  in his Ph.D. thesis (Harvard University, 1966) and related papers in Physical Review Letters and Physical Review.  
In short, he pulled together  two basic pillars of physics: quantum mechanics and general relativity, and realized then that particles can be spontaneously produced by the expansion of the universe or, in general, by  a time-dependent gravitational field.

Around the same time the cosmic microwave background was discovered, changing completely the view of cosmology and reinforcing the Big-Bang theory. In 1992, the COBE satellite detected for the first time small fluctuations in the average temperature of 2.7  Kelvin degrees. This has been confirmed by many other experiments, including the PLANCK satellite. Quantum field theory in curved spacetime and, in particular, cosmological particle creation,  provides 
the  mechanism driving primordial perturbations which seeded the  
tiny fluctuations in temperature observed in the cosmic microwave background.  It 
also helps to explain  the  clumping of matter that gave rise to galaxies and galactic clusters.

In 2014, the  announcement by the BICEP2 team that they had detected the so-called B-modes in the cosmic microwave background (produced by gravitons generated in the very early universe)  greatly excited the physics community. If it had been confirmed, it would have provided  direct evidence of  cosmological particle creation. This may happen in coming years with new observations and experiments. 

Because of the celebration of the centennial, in 2015, of Albert Einstein's theory of General Relativity, the University
of Valencia hosted the Conference ERE2014 (Almost 100 Years after Einstein's Revolution) during
the first week of September. Prof. Parker delivered a  talk  on
the phenomena of gravitational particle creation. J. N-S then took the opportunity to interview him about these topics  from a historical and personal perspective. A very small fraction of the interview was published in {\it Mètode}, the magazine of the University of Valencia. We believe that it will be of interest for young students and researches, and also for senior scientists, to offer the longer and full version of the interview.

\section{Interview: the discovery of  cosmological particle creation}

{\bf  The birth of new fundamental ideas is often  very difficult. In the early sixties, even the novel ideas of spontaneous symmetry breaking of Brout-Englert-Higgs-Guralnik-Hagen-Kibble 
were initially received with skepticism by  some physicists.   In current times the idea of gravitational particle creation seems very natural, but at the time, I guess,  it was not.     How did you experience the initial reactions to the phenomena of cosmological particle creation?}\\

{ Let me start by recalling the context of the time. In 1962, at Harvard when I began my Ph.D.~thesis, I wanted to work at the interface of general relativity and quantum field theory.  I had already studied general relativity quite carefully as an undergraduate before coming to Harvard.  I had the good fortune to 
learn quantum field theory and particle physics at Harvard from Wendell Furry, Roy Glauber (Nobel Prize 2005), Sidney Coleman, Sheldon Glashow (Nobel Prize 1979), and Julian Schwinger (Nobel Prize 1965). 

I wanted to find new consequences of the quantum field theory of elementary particles in the context of Einstein's theory of general relativity.  At the time, I felt that {\em quantizing} the nonlinear gravitational field {\em itself} was so difficult that I would not be able to make significant progress in trying to go beyond the deep work that had already been done in that area.  Nevertheless, I felt that it would be valuable to study quantized elementary particle fields in the curved space-times that were solutions of the nonlinear Einstein gravitational field equations. Luckily, Sidney Coleman agreed to be my thesis advisor on such a project, which was outside the main stream of the time. 

I started by looking for new consequences of quantum field theory in the isotropically expanding cosmological space-times that were solutions of the nonlinear equations of general relativity.}

{ To investigate the creation of particles by the metric of an expanding universe in general relativity, it was first necessary to extend the quantum field theory of elementary particles from the well-established flat Minkowski space-time of special relativity to the context of a classical general relativistic expanding universe with a general expansion scale-factor, $a(t)$. Some examples that I studied were the dust-filled universe, the radiation-dominated universe and the exponentially-expanding universe of deSitter, which at the time had been studied as the steady-state universe, and is now studied as the inflationary universe.  I chose to first consider the spatially-flat universes because I felt that they were the most natural. I developed the quantum field theory by using the simplest generally-covariant extension of the Minkowski space-time field equations, {\em starting with the generally-covariant minimally-coupled scalar-field equation}.

The quantum field theory in these spatially-flat expanding universes could be generated in a straightforward way by simply evolving the well established quantum field theory that was already known in flat Minkowski space-time. The {\it free} (non-interacting) field was chosen, so as not to complicate things with non-gravitational interactions. Then smooth evolution of the free field equation from an initial Minkowski space-time with constant $a(t) = a_1$ to a changing $a(t)$ of an expanding universe (with continuous first and second time derivatives) gave the quantum field theory of the field in the expanding universe, where $a(t)$ was obtained from the Einstein equations of general relativity. It was then straightforward to use the theorem for the most general solution of a second-order differential equation to find that the creation and annihilation operators of the quantized field in the early-time Minkowski space-time each evolved, under the influence of the expansion of the universe, into a superposition of creation and annihilation operators. Since we knew, from the experimentally established interpretation of quantum field theory in flat space-time, how to identify the particles that were created and annihilated during any period when $a(t)$ was constant, we could unambiguously interpret the effect of the expansion of the universe on the particle number, by allowing $a(t)$ to smoothly approach another constant value, $a_2$, at late times. This proved unambiguously that particles were created from the vacuum state of the initial Minkowski space-time having $a(t) = a_1$ because as the evolution showed, there were particles present in the final Minkowski space-time having $a(t) = a_2$.

I completed this part of my Ph.D. Thesis at Harvard by 1964, and went
on to consider spin-1/2 fields, including electrons and positrons in curved
spacetime, and the massless spin-1/2 neutrino field (at that time, it was
not known that there were neutrinos that had mass).  I used the formulation
that had been developed in the 1930's in papers of V. Bargmann and E.
Schrödinger for spin-1/2 particles in curved spacetime. At the suggestion
of my thesis adviser, Sidney Coleman, I wrote an appendix on this
formulation as part of my thesis. As the articles of Bargmann and
Schrödinger were in German, I went through their calculations as best I
could, in my own way, and wrote the appendix in English.

     I finished my full Ph.D. Thesis at Harvard in 1966, during a time when
Sidney Coleman was away in Europe for a couple of years.\\

 {\bf This delayed the publication of your surprising and fantastic result.}\\

As I had to wait until his return before my Thesis defense at Harvard, I
sent a copy of part of my thesis to Professor Bryce S.~DeWitt in early 1966, including the work on particle creation from the vacuum in the expanding universe. DeWitt was one of the leading physicists working on quantum gravity, and I was impressed by an article he had recently published in the Les Houches lectures of 1964.  He was head of the Institute of Field Physics at the University of North Carolina (UNC) at Chapel Hill. He immediately invited me to give a colloquium there and offered me a position there which started in the Fall of 1966.  At the time, he mentioned that I would be taking a position as a postdoc following Peter Higgs. His famous paper on the Higgs boson was written while Higgs was a postdoc at the Institute of Field Physics.  In 2011, the well known physicist, Cecile DeWitt-Morette, who was Bryce DeWitt's wife wrote {\em The Pursuit of Quantum Gravity:Memoirs of Bryce DeWitt from 1946 to 2004}. This is an excellent book, from which I learned a number of interesting things when I reviewed it for the August 2011 issue of Physics Today.  In particular, I learned that Bryce DeWitt had lost much of his funding at about the time I was hired. It was only then that I understood why I was actually hired as an Instructor by UNC with a teaching load of 2 courses per semester, instead of as a postdoc.  As it turned out this teaching was valuable to me, as one of the courses I taught was graduate level quantum mechanics which had been taught regularly by Eugen Merzbacher.  Also valuable were the presence of two prominent physicists from Japan, Ryoyu Utiyama and Tsutomu Imamura, who were visiting the Institute of Field Physics during the time I was there. In the Fall of 1966, I flew back to Harvard  to face my Thesis Defense. I recall being very nervous about it. The examining committee consisted of Sidney Coleman, Sheldon Glashow, and Walter Gilbert (Nobel Prize 1980).  I passed.

My position as Instructor at UNC at Chapel Hill disappeared after 2 years and I was searching for a postdoctoral position in 1968, which was when I was hired as an Assistant Professor by the University of Wisconsin in Milwaukee (UWM).   Again I had a teaching load of two courses each semester, but the first paper on my thesis was published in 1968 in Physical Review Letters. The second was published in 1969 in Physical Review. It is amusing that I found time to write the second paper because my wife and I were stranded in Queens, New York, by a record snow storm that kept us indoors and made it impossible to reach the airport to fly back to Milwaukee for two weeks, during which I wrote the paper. If it were not for the snowstorm, I would have been busy teaching two courses, instead.\\

{\bf Uff!! But, how the paper was received?}\\

As it turned out, I never had a postdoctoral position, but NSF did realize that my work was new and important. 
I did not know it at the time, but my papers were well received, as I later found out.  In 1969, my research proposal was funded by the National Science Foundation; among the projects I suggested in my NSF proposal was to determine the particle creation that would occur when matter collapsed to form a black hole.  In 1970, I received a visit in Milwaukee from Remo Ruffini, an Assistant Professor at Princeton University and Steven A.~Fulling, a graduate student working on a thesis under the direction of Professor Arthur Wightman. In 1971, I was invited to visit Princeton University for the academic year. I have very fond memories of working with the group of Professor John A.~Wheeler, who became interested in my black hole particle creation project.  Also, I was very grateful to Professor Arthur Wightman for meeting with me and inviting me to be the second reader on the Ph.D.~thesis of Steven A.~Fulling.  After completing his Ph.D.~thesis in 1972, I hired Fulling as my first Postdoctoral Fellow with funds from my NSF grant.  Another excellent graduate student at Princeton was Lawrence H.~Ford, who was working with John Wheeler. When Larry Ford completed his Ph.D.~thesis, I hired him as my second postdoc, again using my NSF grant, while Steve Fulling was hired at King's College, London, by Professor Paul Davies. The work being done by my group at UWM stimulated much interest.  Larry Ford was hired at Imperial College London after his postdoc here.  He soon became a faculty member at Tufts University in Boston, and has remained there ever since.  With NSF support, I managed to continue hiring excellent postdocs, and to bring in first class faculty, including John Friedman (a student of S. Chandrasekhar) and Bruce Allen (a student of S. W. Hawking).  Our group continued to grow and do excellent research, eventually becoming the Center for Gravitation, Cosmology and Astrophysics at UWM.  While I was at Princeton in 1971, it became clear that my work had been well received and generated a great deal of interest, particularly in Europe and in the former Soviet Union. It certainly was influential in stimulating further research. 

In the {\it Memoirs} of Andrei Sakharov, he mentions my work of the 1960's on p 261 in Chapter 18: ``Scientific Work of the 1960's". The reference to the Sakharov Memoirs, translated into English from the Russian by Richard Lourie, is: Andrei Sakharov, {\it Memoirs} (Alfred A.~Knopf, Inc., New York, 1990), ISBN 0-394-53740-8. Sakharov was away from the literature when he wrote the {\it Memoirs}, but even a brief mention by such a great man is valuable to me.

Nevertheless, it took time for my work to stimulate much interest in the United States, except for Bryce DeWitt. In fact, I was without a position during the summer of 1968 and chose to use the time to write up some of the work that I had done in my Harvard Ph.D.~thesis. As it turned out, it took some years for my work to be accepted. Many good physicists working on quantum gravity had overlooked particle creation by the classical Einstein gravitational field and did not understand its significance. (One exception was the great Erwin Schr{\"o}dinger who discussed, in 1939, a process in which a propagating wave packet in an expanding universe generated a backward moving wave packet and increased the amplitude of the original wave packet. Unfortunately, he did not use quantum field theory and did not discover the spontaneous creation of pairs of particles from the vacuum in the expanding universe. He regarded the amplification process as ``alarming'' and expected that it would be most significant when the universe reached its maximum expansion and started to contract.)  

By 1972, my work on particle creation by the expanding universe was becoming well known.  For example, in July 1972, I was invited to give a lecture on my work at the Aspen Center for Physics.  Also notable, was a series of invited lectures on particle creation by the expansion of the universe and by black holes that I gave in 1975 at the Second Latin American Symposium on Relativity and Gravitation held in Caracas, Venezuela, chaired by Carlos Aragone at Universidad Simon Bolivar.  Among the other speakers were M.~Kaku, A.~Papapetrou, T.~Regge, K.~Thorne, C.~Teitelboim, P.~van Niewenhuizen, L.~Witten, and B.~Zumino.   I also was invited to lecture on my work at an NSF funded conference chaired by Louis Witten and F. Paul Esposito at the University of Cincinnati in 1976, published by Plenum Press, New York, in 1977, ISBN 0-306-31022-8.  In July 1978, I lectured on {\em Aspects of Quantum Field Theory in Curved Space-Time: Effective Action and Energy-Momentum Tensor} at the 1978 Carg{`e}se Summer Institute on Recent Developments in Gravitation, edited by Maurice L{'e}vy and S.~Deser, ISBN 0-306-40198-3. Among the other notable lecturers there were S.~W.~Hawking, B.~S.~DeWitt and G. `t Hooft. I was pleased to receive a postcard from John A.~Wheeler complimenting me on the lecture I gave at Carg{`e}se.

In my paper of 1968 in Physical Review Letters, I stated in the conclusion section that the reaction back of the particles created by the gravitational field in the early expanding universe would be large and would have to be taken into account in discussing the particle creation in the early universe. In 1973, Fulling and I published a paper in Physical Review, in which we numerically integrated the Einstein equations including the reaction back of the matter, taken to be in a Bose-Einstein condensed state. We showed that a "bounce" could occur, in which the universe went from a collapse to an expansion. This showed explicitly that it was possible for a plausible quantum state of matter to elude the cosmological singularity that would have followed from the classical matter ``energy-conditions,'' with interesting consequences for the universe.\\

 {\bf Let us go back to the question of black holes and particle creation. When did you first think about that?}\\

Before leaving Harvard, I suggested to my advisor that I believed it would be interesting to calculate the particle creation by the gravitational field of matter collapsing to form a black hole. He suggested that I write to John Wheeler at Princeton when applying for a postdoc position. I did that but the reply said that no postdoc position was available, but if I had funds of my own, I could come to Princeton to work on the project. I also mentioned the idea to Bryce DeWitt when I was an instructor at UNC. He agreed that the particle creation should occur, but did not express an interest in working on such a project with me. Luckily, NSF expressed an interest when I applied for a grant in 1968 and included in my proposal the idea of calculating the particle creation by a black hole. As I mentioned above, I did go to Princeton and the Wheeler group in 1971, using my NSF funds, and suggested doing the calculation to several people. However, my stay at Princeton University ended in August 1972, when I had to return to my position at UWM, with a heavy teaching load. I was working on the project of particle creation in the formation of a black hole, when I was shown Hawking's beautiful paper of 1974 on that topic and immediately recognized that Hawking had solved the black hole particle creation problem in a beautiful and insightful way. 

To my regret, in his paper Hawking did not cite my earlier papers of 1968 and 1969 on particle creation in an expanding universe, in which I showed that the expansion of the universe caused creation operators to evolve into superpositions of creation and annihilation operators. This corresponds to the creation of particle-antiparticle pairs, and the transformation is known as a Bogoliubov transformation. I had discussed this transformation in detail in my Physical Review paper of 1969 on particle creation by the expanding universe. The same type of superposition of creation and annihilation operators was found by Hawking to occur when a dust cloud collapses to form a black hole.  I wrote a note to Hawking and he did add a reference to my work in a later review article on particle creation by black holes that he wrote.

In a 1973 paper published in Physical Review, Steven Fullling showed that a similar superposition of creation and annihilation operators occurs in the Rindler coordinate system corresponding to a set of accelerated observers in a region of Minkowski space. This work of Fulling was referenced in the original 1974 Hawking paper and serves as a link back to my 1968 and 1969 papers on particle creation by an expanding universe.

 Despite the fact that the link to my work on particle creation by expanding universes was ignored, I did soon make a significant contribution in 1975 to particle creation by black holes by working out the detailed probability distribution of the particles created by a black hole. This appeared in 1975 in Physical Review.  The method that I used was developed in my Ph.D.~thesis to calculate the probability distribution of pairs of particles created by the expanding universe.  Around that time, as already mentioned a few paragraphs above, I also gave some lecture series, published in various Proceedings, in which I discussed and explained aspects of particle creation by expanding universes and by black holes.   These lectures were found valuable by many students and researchers around the world.}\\

{\bf The phenomena of gravitational particle creation has been  acquiring different faces. Using your formalism   S. W. Hawking realized in 1974 that black holes also create particles, in a way deeply connected with thermodynamics. How did it influence the subsequent research in the field? }\\

{ Hawking's beautiful result was very influential. It convinced everyone that the second law of thermodynamics was valid for systems that included black holes. This revealed a deep connection between thermodynamics and general relativity. It was very gratifying to me that particle creation by a black hole played such a central role in showing the consistency of thermodynamics in the presence of black holes. The fact that the area of a black hole was related to its temperature and to the actual entropy of a black hole had already been argued in the published literature by Jacob Bekenstein, who by the way, was a student of John Wheeler during the time that I was at Princeton. Bekenstein argued in his Ph.D.~thesis that the area of a black hole was proportional to the  entropy of the black hole as it appeared in an equation that emerged from general relativity for a system containing black holes. However, it had been argued in the literature that this could not be a correct interpretation because black holes seemed to absorb the entropy of bodies falling into them, thus violating the second law of thermodynamics. However, the particle creation process revealed the precise temperature of the radiation emitted by the black hole, which had a thermal spectrum. Furthermore, it showed that general relativity remained consistent with the second law of thermodynamics. These results revealed a deep consistency between general relativity, including its prediction of black holes, and quantum field theory in the curved spacetime revealed by general relativity. Already in my 1969 paper in Physical Review, I pointed out that the process of particle creation by the expansion of the universe revealed a deep consistency between the macro scales dealt with by classical general relativity and the small scales normally dealt with by quantum field theory. 

What I pointed out in my paper of 1969 was that the equations of general relativity were consistent with the particle creation results obtained by using quantum field theory in the expanding universes of general relativity. I illustrated this by showing that the creation of massless scalar particles went to $0$ in the classical general relativistic expanding universe that was dominated by massless particles (often called the radiation-dominated universe), and that the creation of massive scalar particles went rapidly to $0$ (as a function of increasing mass) in the classical general relativistic universe that was dominated by massive particles (often called the ``dust filled'' universe). As I stated in my 1969 paper, this revealed a ``deep consistency" between general relativity and quantum field theory. One could think of these universes and their matter content as being examples of thermodynamic equilibrium "stationary" states in which the matter content is consistent with the spacetime generated through the Einstein field equations.}\\

{\bf Soon after these works on black holes you were involved in the study of graviton production (in collaboration with L. Ford, 1977).  The (isotropic) expansion of the universe is not able to create photons or massless fermions. Were you surprised to see that the  expanding universe was actually able to create (massless) gravitons? Did you think that  sophisticated experiments (like those carried out in current times) may even detect the effects of those gravitons?}\\

{ Let me first fill in some interesting background relevant to your question.  As I was working on my Ph.D.~thesis at Harvard, after proving that {\it minimally-coupled} scalar particles were created by the isotropically expanding universe, I also determined the form of the scalar field equation for which no particles would be created in such a universe for an arbitrary smooth isotropic expansion. The mass of the particle had to be exactly $0$ and there were terms involving the first and second time derivatives of the scale factor $a(t)$ in the equation. These terms were a multiple of the Ricci scalar curvature in such expanding universes.

About a year later, in 1964, I was reading {\it Relativity, Groups and Topology}, edited by C.~DeWitt and B.~S.~DeWitt. In this collection of lectures given in 1964 at the Les Houches summer school, I stumbled upon a brief 2 page article by Roger Penrose in which he wrote down the {\it conformally-invariant} wave equation for the classical massless scalar field in a {\em general} 4-dimensional curved spacetime. The equation involved the Ricci scalar curvature of the spacetime.  When I compared it with the particular scalar field equation that I had already shown to have $0$-particle creation for any smoothly and isotropically expanding universe, I was surprised to see that my equation was the same as the conformally-invariant scalar field equation given by Penrose, when the latter was written for a general smoothly and isotropically expanding universe. 

Furthermore, I had already shown that the quantized massless neutrino field equation of Pauli, which already was written in terms of 2-component spinors, created no massless neutrinos in any smoothly and isotropically expanding universe. (In the 1960's neutrinos were generally believed to have $0$-mass.)

In his brief article, Penrose had also written down, in 2-component spinor form, the conformally invariant field equations for all fields of half-integer and integer spin.  Among these conformally-invariant field equations was the Pauli 2-component spinor neutrino equation for a massless spin-1/2 neutrino. It seemed obvious to me that there must be a deep connection between particle creation and conformal invariance in such universes. 

The {\em new} result that I proved in my thesis was that for all spins there was no creation of massless particles satisfying the Penrose conformally-invariant field equations in any isotropically and smoothly expanding universe. The proof made use of the fact that such universes were {\em conformally flat}. This result was published in my Physical Review papers of 1968 and 1969. This was the {\it first} published work showing the connection between conformal invariance and particle creation in isotropically expanding universes.  

Among the conformally-invariant equations given by Penrose was the spin-2 massless field satisfying a conformally-invariant field equation. At the time (about 1964), I thought this implied that in the isotropically expanding universes that I was considering, which were known to be conformally flat, there would be no creation of gravitons, although in anisotropically expanding universes gravitons would still be created. In my published papers of 1968 and 1969, I had stated that there would be no creation of linearized {\em conformally-invariant} gravitons in the isotropically expanding universes. When I visited John A. Wheeler's group at Princeton University in 1971, Karel Kucha{\~r} explained to me that if one used a conformally invariant gravitational field equation, then it had been shown that the graviton could not couple to matter, as it {\em does} in the Einstein gravitational field equations.  As a consequence, it was clear that the linearized Einstein equation could not be the the same as the conformally invariant spin-2 wave equation.

Thus, to answer your question, I was {\em not} surprised by the fact that linearized Einstein gravitons would be created by isotropically expanding universes. However, after returning to UWM from Princeton in 1972 I was occupied with teaching some large undergraduate classes and with trying to solve the black hole particle creation problem. If I had known about the Lifshitz gauge of 1948, in which Lifshitz had shown that the linearized Einstein equation reduces to two minimally-coupled massless scalar field equations, it would have been obvious from my work on minimally-coupled scalar particles, that Einstein gravitons would be created in isotropically expanding universes.

A number of years later, in 1975 Leonid Grishchuk, then at Moscow State University, published a paper in JETP pointing out that E.~M.~Lifshitz had proved in 1946 that, in the so-called Lifshitz gauge, the linearized Einstein equation reduces to a spin-2 field equation; and that each component of the spin-2 graviton field, in that gauge satisfies the {\em minimally-coupled} scalar field equation!  I had already shown that those minimally-coupled scalar particles were indeed created in an isotropically expanding universe, so Lifshitz work of 1946 together with my results published in 1968 and 1969 implied that Einstein gravitons (i.e., quantized gravitational wave perturbations) can be created by isotropically expanding universes.
 
In 1977, Lawrence H.~Ford and I wrote a very nice paper in Physical Review on graviton creation.  In our paper we included the spatially-curved isotropically expanding universes, in addition to the spatially-flat expanding universes.  We explicitly quantized the linearized Einstein graviton field and evolved it in the background classical isotropically expanding universes.  We showed that quantized Einstein gravitons of spin-2 were created by the same mechanism that I had used in the 1960's to show that minimally-coupled quantized scalar particles were created, but now applied to each of the two spin components of the massless spin-2 graviton field. We also calculated explicit results for power-law expansions in the spatially-flat case to show that there were no infrared divergences. For the spatially-flat exponentially expanding (inflationary) universe, it would have been straightforward to solve our equations in terms of Hankel functions and to calculate the properties of the created gravitons. The exponentially expanding universe did not become of great interest until the 1980's when the advantages of inflation were pointed out in papers of Guth, Steinhardt and others.\\

 {\bf Did you think that  sophisticated experiments (like those of BICEP2 and the Planck satellite) may even detect the effects of those gravitons?}\\

I must admit that I was excited when I learned that the effect of the created gravitational waves on the polarization of the CMB was actually within the range of current instrumentation!  It is now up to the observers to separate the effect of background from the effect of gravitational waves coming from the very early universe.} \\

 {\bf In your first works you emphasized that particle creation should be very relevant for the very early universe. This prophecy fitted like a glove with the  
proposal of cosmic inflation by A. Guth in 1980.  Could you explain how this happens?}\\

{ First of all, when I was working on my Ph.D.~thesis, I calculated the number density of minimally-coupled scalar particles, and also fermions, created by typical expanding universes and found that the density of created particles was higher if one included an early period of rapid expansion of the universe. This was in fact fairly obvious because the number of created particles typically depended on terms like $({\dot{a}(t)/a(t))^2}$ and ${\ddot{a}(t)/a(t)}$ that are relatively large in the early universe.  I did not discuss this in my thesis because my advisor wanted me to concentrate in my thesis on the creation of particles by the {\em present} expansion of the universe, although I did tell him that it would be most significant in the very early stages of the expanding universe.  In 1975, I published a paper in {\em Nature} on the creation of a thermal spectrum of particles by a class of early expanding universes. These expanding universes had several adjustable parameters and the density of created particles could be obtained analytically.  This set of universes included the exponentially expanding universe. However, at the time, inspired by Hawking's black hole result, my emphasis at the time was on the possibility of creating a thermal spectrum directly from the expansion of the universe. The temperature of the created thermal radiation depended on the values of the adjustable parameters. In the paper, I pointed out that such a process could account for the temperature and entropy of the early universe as required by the consistency of the Einstein equations governing the expansion of the universe.  The main difference from the present inflationary model is that my class of models started from an initial Minkowski spacetime with an initial Minkowski vacuum state and evolved from there.

In 2008 I and my Ph.D. thesis student, Matthew Glenz, generalized the work that I had published in {\em Nature} in 1975 by smoothly joining the model to an exponentially expanding stage of the universe. The solutions for the expansion of the universe and the Bogoliubov coefficients that governed the creation of particles were known analytically in terms of special functions. Our paper on this work was published in Physical Review D.

The joining between the initial stage and the exponential stage of the expansion was continuous in the scale factor $a(t)$ and in its first and second time derivatives, that is, $a(t)$ was a $C(2)$ function, as that was necessary to avoid discontinuities and ultraviolet divergences in the rate of creation of particles by the expansion of the universe, as I had found long ago in my Ph.D.~thesis. What Glenz and I did was to evolve the known special function solutions governing the creation of particles by this expanding universe to over 200 place accuracy to check that the constraint conditions satisfied by the coefficients of the Bogoliubov transformation of creation and annihilation operators were satisfied to enough accuracy (over 200 digits) to accommodate more than 60 e-foldings of exponential expansion. To interpret the results, we also did a $C(2)$ joining of the exponentially expanding $a(t)$ to another function that smoothly approached a late-time Minkowski space (we could also have joined to a late-time slowly expanding universe without significantly changing our conclusions). The main thing that I want to mention is that in our model the initial Minkowski vacuum state evolved into the vacuum state (the Bunch-Davies vacuum proposed in 1979) of the inflationary universe models. It is quite natural that the initial Minkowski space vacuum state, which is defined by the $10$ Poincar{\'e} symmetries, evolved into the deSitter space vacuum, which is defined by the $10$ symmetries of deSitter space.  In the lowest frequency modes having wavelengths that were already outside the inflationary Hubble scale at the beginning of the inflationary stage, the wavelengths remained larger than the inflationary Hubble horizon length scale throughout the period of inflation and the corresponding wavelengths today are larger than our present observable universe. Thus, I do not believe that one can find any observational differences between the predictions of the inflationary models that start with the Bunch-Davies de Sitter vacuum state, or those that start with the Minkowski vacuum state, as I had considered originally in my thesis back in the 1960's.

In fact, our universe may well have started as a spontaneous fluctuation in a flat Minkowski spacetime in the global Poincare vacuum undergoing the natural vacuum fluctuations.  The chance of such a fluctuation becoming like our present universe is greatly favored if it underwent a rapid early stage of exponential inflationary expansion. Now one may ask how the particle creation by the expansion of the universe  may lead to a natural reheating process that ends the exponential expansion and turns it into the standard radiation-dominated expanding universe.  Some of the scenarios already suggested for reheating in the standard inflationary model would also work in the spontaneous fluctuation inflationary models that I described above. One should also not forget that it has been shown that the particle creation process would rapidly bring about isotropic expansion of the universe, if the rapid initial expansion of the universe were anisotropic.

In any event, your question asked how my early suggestion seems to ``fit like a glove" with the proposal of cosmic inflation by A. Guth in 1980. I think that I have given a fairly satisfactory answer based on the framework of quantum field theory in expanding universes that I developed in the 1960's.  }\\

{\bf  After the discovering of the ``Higgs'' particle, do you think it is fair to say that the graviton  has been converted into physics most-wanted particle (perhaps sharing the place with the super-particles)? }\\

{  Yes, indeed. The {\em graviton} is the massless spin-2 particle that carries a quantum of energy proportional to the frequency of a gravitational wave. (Similarly, the photon is the massless spin-1 particle that carries a quantum of energy proportional to the frequency of an electromagnetic wave.) The loss of energy through the emission of gravitational waves is thought to have been measured from very accurate timing of the orbital period of the Hulse-Taylor binary pulsar system. Although gravitational waves themselves have not yet been {\em directly} measured by LIGO or other gravitational wave detectors on the earth, it is very likely that they will be detected within a few years, as the sensitivity of gravitational wave detectors on earth and in space becomes greater. Until recently, the detection of {\em gravitons}, the quanta that are carried by gravitational waves, seemed outside the range of observation, because their interaction with matter is exceedingly small. That has all been changed by the possibility that we are detecting gravitons created by the very rapid early expansion of the universe.

The polarization pattern of the cosmic microwave background (CMB) is now being measured with extremely sensitive detectors. Using these detectors, we may be able to measure polarization patterns in the CMB that are the signature produced by gravitons that were created in the earliest stages of the expanding universe. Only that source would be powerful enough to create gravitons with sufficient intensity for their effects to be seen in the polarization patterns that they induce in the CMB radiation.

The BICEP2 collaboration has used such detectors at the South Pole to detect polarization patterns in the CMB that appear to be caused by gravitons that were created in the earliest stages of the expanding universe. The Planck satellite has detected related polarization patterns that may have been produced by the effect of dust on the CMB radiation. Currently, the two groups are combining and analyzing their respective measurements in order to arrive at a conclusive result concerning the possible measurement of gravitons coming from the earliest stage of the expanding universe.

Once the effect of dust on the CMB polarization patterns, as already measured by the Planck satellite, has been carefully taken into account, we should be able to tell, with considerable certainly, if the polarization measured by the BICEP2 collaboration at the south pole is the result of quantized gravitons that were created by the expansion of the very early universe. The detection of the signature of these gravitons in the polarization pattern of the CMB radiation certainly would be one of the most remarkable achievements of our time! However, this may need  a
new generation of detectors, with more sensitivity.  And may take several
more years.

LIGO has recently detected gravitational waves from the inspiral and merger
of several black hole binaries having masses of about 20 or 30 solar
masses, but it still remains to directly detect the gravitons associated
with gravitational waves.  So the graviton still remains the ``most wanted
particle''.}\\

{\bf Obviously, you could have been  professor at Harvard or at any other of the most coveted universities. Why did you decide to stay at the University of  Wisconsin-Milwaukee?}\\

I have always had a clear view on this. We had no intention of uprooting our children to push my scientific position. I could have been very naive, but I  
do not regret. \\

{\bf A more personal question. You play piano and also collect paintings of the 17th century. How did your humanistic interests influence  your work and  your view of the universe? }\\

{ Even as a young boy I was greatly fascinated by art and music. I studied the drawings of artists such as Leonardo da Vinci and tried to copy them.  I enjoyed making pencil and charcoal drawings, water colors, and oil paintings.  As a young teen ager, I bought some sculpture tools and some marble and tried to sculpt a block of marble. However, it was literally too hard for me.  I took piano lessons from the age of about 8 through 11, and then studied piano again as a teen ager because I found classical music so beautiful.  At about the age of 12, I became greatly interested in science and started reading about atoms and nuclei with great interest. In high school, I became very interested in chemistry, physics, and biology, including the genetics of fruit flies.  My interest in science and physics were like my interests in art and music, in that I was greatly influenced by their depth and beauty. Even later, I never felt that I was doing physics as a vocation, but rather as an art of great depth and beauty. It was only when I realized that one had to support a family that I focused on publishing in order to get promoted to a secure position. I still feel like an artist with regard to doing research and I continue to do it, as best I can, although I am retired.}






\begin{thebibliography}{99}

\nonfrenchspacing


\bibitem{parker66} Leonard Parker, {\it The creation of particles in an expanding universe}, Ph.D. thesis, Harvard University (1966).  To order a copy of this thesis,
use the internet to go to Dissexpress.umi.com and enter the Publication
Number  7331244.

\bibitem{parker68} L. Parker, {\it Phys.Rev.Lett.} {\bf 21} 562 (1968).
\bibitem{parker69a} L. Parker, {\it Phys. Rev.} {\bf 183}, 1057(1969).
\bibitem{parker69b} L. Parker, {\it Phys. Rev. D} {\bf 3}, 346 (1971).
\bibitem{parker-fulling} L. Parker and S. Fulling,  {\it Phys.Rev. D} {\bf 7}, 2357 (1973).

\bibitem {Hawking} S.W. Hawking, {\it Nature} {\bf 248}, 30 (1974).



\bibitem{Grishchuk} L.P. Grishchuk, {\it Sov. Phys. JETP} {\bf 40}, 409 (1975).

\bibitem{Parker} L. Parker, {\it Nature} {\bf 261}, 20 (1976).

\bibitem{ford-parker} L. H. Ford and L. Parker, {\it Phys. Rev. D} {\bf 16}, 1601 (1977).

\bibitem{guth} A. H. Guth {\it Phys.Rev. D} {\bf 23}, 347 (1981).

\bibitem{glenz-parker} M. M. Glenz and L. Parker, {\it Phys.Rev. D} {\bf 80}, 063534 (2009).

\bibitem{navarro-salas} J. Navarro-Salas, {\it Mètode} {\bf 83}, 10 (2014).
\bibitem{parker2015} L. Parker,  {\it Creation of quantized particles, gravitons and scalar perturbations by the expanding universe}, arXiv:1503.00359.

\end{thebibliography}
\end{document}